\documentclass[twocolumn,
prd,
aps,
superscriptaddress,
preprintnumbers,
tightenlines,
showpacs,
nofootinbib,
eqsecnum,
amsfonts,
amsmath]{revtex4-1}

\usepackage{chronology}
\usepackage{tikz}
\usetikzlibrary{decorations.pathreplacing}
\usepackage{graphicx}
\usepackage{dcolumn}
\usepackage{bm}
\usepackage{hyperref}
\usepackage{lineno}
\usepackage{ulem}
\usepackage[bottom]{footmisc}
\usepackage{footnote}

\newcommand{\mf}[1]{{#1}}

\begin{document}


\title{Modeling non stationary noise in pulsar timing array data analysis}

\author{Mikel Falxa}
\email{mikel.falxa@unimib.it}
\affiliation{Laboratoire de Physique et Chimie de l'Environnement et de l'Espace, Universit\'e d’Orl\'eans/CNRS, 45071 Orl\'eans Cedex 02, France}
 \affiliation{Dipartimento di Fisica “G. Occhialini", Università degli Studi di Milano-Bicocca, Piazza della Scienza 3, I-20126 Milano, Italy}

\author{J. Antoniadis}
\affiliation{FORTH Institute of Astrophysics, N. Plastira 100, 70013, Heraklion, Greece}
\affiliation{Max-Planck-Institut f\"ur Radioastronomie, Auf dem H\"ugel 69, D-53121 Bonn, Germany}

\author{D. J. Champion}
\affiliation{Max-Planck-Institut f\"ur Radioastronomie, Auf dem H\"ugel 69, D-53121 Bonn, Germany}

\author{I. Cognard}
\affiliation{Laboratoire de Physique et Chimie de l'Environnement et de l'Espace, Universit\'e d’Orl\'eans/CNRS, 45071 Orl\'eans Cedex 02, France}
\affiliation{Observatoire Radioastronomique de Nan\c{c}ay, Observatoire de Paris, Universit\'e PSL, Universit\'e d’Orléans, CNRS, 18330 Nan\c{c}ay, France}

\author{G. Desvignes}
\affiliation{Max-Planck-Institut f\"ur Radioastronomie, Auf dem H\"ugel 69, D-53121 Bonn, Germany}

\author{L. Guillemot}
\affiliation{Laboratoire de Physique et Chimie de l'Environnement et de l'Espace, Universit\'e d’Orl\'eans/CNRS, 45071 Orl\'eans Cedex 02, France}
\affiliation{Observatoire Radioastronomique de Nan\c{c}ay, Observatoire de Paris, Universit\'e PSL, Universit\'e d’Orléans, CNRS, 18330 Nan\c{c}ay, France}

\author{H. Hu}
\affiliation{Max-Planck-Institut f\"ur Radioastronomie, Auf dem H\"ugel 69, D-53121 Bonn, Germany}

\author{G. Janssen}
\affiliation{ASTRON, Netherlands Institute for Radio Astronomy, Oude Hoogeveensedijk 4, 7991 PD, Dwingeloo, The Netherlands}
\affiliation{Department of Astrophysics/IMAPP, Radboud University Nijmegen, P.O. Box 9010, 6500 GL Nijmegen, The Netherlands}

\author{J. Jawor}
\affiliation{Max-Planck-Institut f\"ur Radioastronomie, Auf dem H\"ugel 69, D-53121 Bonn, Germany}

\author{R. Karuppusamy}
\affiliation{Max-Planck-Institut f\"ur Radioastronomie, Auf dem H\"ugel 69, D-53121 Bonn, Germany}

\author{M. J. Keith}
\affiliation{Jodrell Bank Centre for Astrophysics, Department of Physics and Astronomy, University of Manchester, Manchester M13 9PL, U}

\author{M. Kramer}
\affiliation{Max-Planck-Institut f\"ur Radioastronomie, Auf dem H\"ugel 69, D-53121 Bonn, Germany}

\author{K. Lackeos}
\affiliation{Max-Planck-Institut f\"ur Radioastronomie, Auf dem H\"ugel 69, D-53121 Bonn, Germany}

\author{K. Liu}
\affiliation{Shanghai Astronomical Observatory, Chinese Academy of Sciences, 80 Nandan Road, Shanghai 200030, China}
\affiliation{Max-Planck-Institut f\"ur Radioastronomie, Auf dem H\"ugel 69, D-53121 Bonn, Germany}

\author{J. W. McKee}
\affiliation{E. A. Milne Centre for Astrophysics, University of Hull, Cottingham Road, Kingston-upon-Hull, HU6 7RX, UK}
\affiliation{Centre of Excellence for Data Science, Artificial Intelligence and Modelling (DAIM), University of Hull, Cottingham Road, Kingston-upon-Hull, HU6 7RX, UK}

\author{D. Perrodin}
\affiliation{INAF - Osservatorio Astronomico di Cagliari, via della Scienza 5, 09047 Selargius (CA), Italy}

\author{S. A. Sanidas}
\affiliation{Jodrell Bank Centre for Astrophysics, Department of Physics and Astronomy, University of Manchester, Manchester, M13 9PL}

\author{G. M. Shaifullah}
\affiliation{Dipartimento di Fisica "G. Occhialini", Università degli Studi di Milano-Bicocca, Piazza della Scienza 3, I-20126 Milano, Italy}
\affiliation{INFN, Sezione di Milano-Bicocca, Piazza della Scienza 3, I-20126 Milano, Italy}
\affiliation{INAF - Osservatorio Astronomico di Cagliari, via della Scienza 5, 09047 Selargius (CA), Italy}

\author{G. Theureau}
\affiliation{Laboratoire de Physique et Chimie de l'Environnement et de l'Espace, Universit\'e d’Orl\'eans/CNRS, 45071 Orl\'eans Cedex 02, France}
\affiliation{Observatoire Radioastronomique de Nan\c{c}ay, Observatoire de Paris, Universit\'e PSL, Universit\'e d’Orléans, CNRS, 18330 Nan\c{c}ay, France}
\affiliation{Laboratoire Univers et Th\'eories LUTh, Observatoire de Paris, Universit\'e PSL, CNRS, Universit\'e de Paris, 92190 Meudon, France}

 

\date{\today}

\begin{abstract}
Pulsar Timing Array (PTA) collaborations recently reported evidence for the presence of a gravitational wave background (GWB) in their datasets. The main candidate that is expected to produce such a GWB is the population of supermassive black hole binaries (SMBHB). Some analyses showed that the recovered signal may exhibit time-dependent properties, i.e. non-stationarity. In this paper, we propose an approximated non-stationary Gaussian process (GP) model obtained from the perturbation of stationary processes. The presented method is applied to the second data release of the European pulsar timing array to search for non-stationary features in the GWB. We analyzed the data in different time slices and showed that the inferred properties of the GWB evolve with time. We find no evidence for such non-stationary behavior and the Bayes factor in favor of the latter is $\mathcal{B}^{NS}_{S} = 1.5$. We argue that the evolution of the GWB properties most likely comes from the \mf{improvement of the observation cadence} with time and \mf{better} characterization of the noise of individual pulsars. Such non-stationary GWB could also be produced by the leakage of non-stationary features in the noise of individual pulsars or by the presence of an eccentric single source.

\end{abstract}

\maketitle

\section{Introduction}

Supermassive black holes (SMBHs) are incredibly massive, over a million times heavier than the sun, and are typically located at the centers of galaxies. When these black holes come together, they form binary systems that emit strong gravitational waves as they approach each other \cite{Sesana_2008}. The combined effect of many such binaries produces a continuous and stationary noise signal in the nanohertz frequency band \cite{phinney2001practical}, known as the gravitational wave background (GWB). Pulsar timing array (PTA) collaborations search for the induced effect of this GWB in the timing measurements of millisecond pulsars. They specifically target the characteristic spatial correlations it would exhibit between pairs of pulsars, known as the Hellings-Downs (HD) correlation pattern \cite{Allen_2023}. Recent studies have shown strong evidence for the presence of such a signal \cite{Agazie_2023, ppta_dr3, wm3}. However, it does not exactly match what was expected for a population of supermassive black hole binaries (SMBHB) in circular orbits. In particular, it was observed that the inferred GWB properties differed when datasets of different lengths were analyzed. Several hypotheses have been proposed to explain this discrepancy, one of which is the possible non-stationarity of the GWB.

The European Pulsar Timing Array (EPTA) collaboration has released its second dataset, combining the timing data of 25 pulsars over a span of up to 24.7 years of observation \footnote{The EPTA data is publicly available at \url{https://epta.pages.in2p3.fr/epta-dr2/}}. Several radio telescopes were used to collect the data: the Effelsberg Radio Telescope in Germany, the Lovell Telescope in the UK, the Nançay Radio Telescope in France, the Westerbork Synthesis Radio Telescope in the Netherlands, the Sardinia Telescope in Italy, and the Large European Array for Pulsars. In their analysis \cite{wm3}, the EPTA searches for the presence of a GWB in the dataset by evaluating the Bayes factor, comparing common HD correlated noise against common uncorrelated noise denoted as $\mathcal{B}^{HD}_{CURN}$. The significance and properties of the GWB vary notably across the dataset. When analyzing the entire dataset, the Bayes factor is only $\mathcal{B}^{HD}_{CURN}=4$, but it is shown that using only the latest 10.33 years of data, the Bayes factor rises to $\mathcal{B}^{HD}_{CURN}=60$. The justification for trimming the dataset is the evolution of data quality. Legacy data have a worse cadence of observation and fewer multi-frequency band observations \cite{wm1}. These two qualities are crucial for performing a good characterization of all sources of noise present in the timing data, essential for an accurate observation of the GWB \cite{wm2}. \mf{Radio observatories are equipped with receiver-backend systems where the receiver collects the raw radio data that goes through a real-time processing stage in the backend systems, providing analyzable data \citep{wm1}}. The latest 10.33 years of data were obtained from new-generation backends with greatly improved observing abilities. However, the evolution of properties of the observed signal was not only reported by the EPTA collaboration. The Parkes Pulsar Timing Array (PPTA) performed a time-slice analysis where their dataset was analyzed in slices of 3 or 9 years and showed that the amplitude of the signal varied with time \cite{ppta_dr3}. The reason behind this time-evolving property remains unexplained as the GWB is expected to be stationary \cite{phinney2001practical, Sesana_2008}. One advanced hypothesis is non-stationarity of noise present in the pulsars themselves that would contaminate our estimate of the signal. The two dominant sources of time-correlated noise in pulsar timing are the red noise (RN), due to stochastic variation of the pulsar's spin rate, and the dispersion measure (DM) noise, due to the fluctuation of the electron density in the interstellar medium during the path that the pulses take to Earth \cite{wm2} (this noise has an amplitude proportional to $\nu^{-2}$ with $\nu$ being the incoming photon's frequency). The noise contributions are typically assumed to be stationary. Other potential sources of non-stationarity of astrophysical origin are eccentric SMBHBs. They produce gravitational waves (GW) with significantly varying frequency content within one orbital period of the binary \cite{Taylor_ecc_2016, Susobhanan_2020}. The presence of one or more eccentric sources could influence a stationary GWB and make it non-stationary.

In this paper, we propose a way to model non-stationary noise. Previous works explored the idea of modeling non-stationary signals in GW analysis using wavelets \cite{Ellis_2016, cornish_wdm}. Here, we develop a model which is a perturbation of the Gaussian process (GP) methods that are commonly used in PTA data analysis \cite{new_advances}. The main motivation behind this model is to evaluate the significance of non-stationary features in noise through the evaluation of the Bayes factor by allowing time dependence of the noise spectrum properties. Strong variations in time result in strong correlations between different frequencies of the spectrum that are not accounted for by ordinary Fourier decomposition \cite{cornish_wdm}. In general, the data are cut into smaller intervals where the noise is considered approximately stationary, and a Fourier transform is applied in each window to monitor the evolution of the frequency content. The assumption of local stationarity presupposes that the properties of the spectrum evolve slowly in the analyzed interval \cite{mallat}.

This paper is organized as follows. In the first section, we will present a non-stationary GP model. \mf{The ability of this model to correctly infer parameters is then tested on mock datasets in which non-stationary noise is injected.} Finally, we will search for the possible presence of a non-stationary GWB in the EPTA dataset and argue that nearby eccentric SMBH binaries could introduce non-stationary features in the GWB.

\section{Model}

\subsection{Stationary gaussian process}

A stationary process is a stochastic process whose properties, typically its mean and variance, remain constant over time. In practice, we often encounter what we refer to as weak-sense stationary processes, where the mean and correlation functions remain invariant under shifts in time, i.e., they are time-translation symmetric \cite{papoulis02}. Given the one-sided power spectral density (PSD) $S_0(f)$ of the process, the correlation function $C(t,t')$ can be expressed using the Wiener-Khinchine integral :

\begin{equation}
    C(t, t') = \int_0 ^\infty df S_0(f) \cos(2\pi f (t-t'))
\label{eq:wiener_khinchin}
\end{equation}
which is a function of the time difference $t-t'$.

Alternatively, we can define a zero-mean stationary Gaussian Process (GP) expressed as a finite Fourier sum with normally distributed coefficients \cite{new_advances, Rasmussen2004}

\begin{equation}
\begin{aligned}
    n_0(t) & = \sum_{i=1} ^N X_i \sin(2\pi f_i t) + Y_i \cos(2\pi f_i t)\\
    & = \sum_{i=1}^N \vec{\omega}_i \cdot \vec{\phi}_i (t)
\end{aligned}
\label{eq:time_domain_stat_gp}
\end{equation}
where $\vec{\phi}_i(t) = [\sin(2\pi f_i t), \cos(2\pi f_i t)]$ and $\vec{\omega}_i = [X_i, Y_i]$ with $X_i, Y_i \sim \mathcal{N}(0, \sigma_i^2)$.

The basis $\vec{\phi}_i(t)$ of cosine and sine functions is evaluated at a discrete set of frequencies $f_i$. For stationary signals, the random weights $\vec{\omega}_i$ are uncorrelated because the underlying Fourier basis of the process is orthogonal. Thus, using $\langle \vec{\omega}_i, \vec{\omega}_j \rangle = \langle \omega^2 \rangle {ij} = \sigma_i ^2 I_2 \delta{ij}$ (where $\langle a, b \rangle$ denotes the \mf{ensemble averaged} inner product between $a$ and $b$, and $I_2$ is the identity matrix of rank 2), we have:

\begin{equation}
\begin{aligned}
    C(t, t') & = \langle n(t), n(t') \rangle \\
    & = \sum_{ij} \vec{\phi}_i^\top (t) \langle \omega^2 \rangle _{ij} \vec{\phi}_j(t') \\
    & \equiv \sum_i \sigma_i^2 \cos(2\pi f_i (t-t'))
\label{eq:covariance_stat}
\end{aligned}
\end{equation}
which is an approximation of the integral in \autoref{eq:wiener_khinchin} for $\sigma_i^2 = S_0(f_i) \Delta f_i$ with $\Delta f_i = f_{i+1}-f_i$ \cite{lowrank}.

\subsection{Non-stationary gaussian process}
\label{subsec:ns_gp}

By definition, a non-stationary process exhibits spectral properties that vary with time \cite{priestley, Hong_2023}. Let us consider a time-dependent evolutionary power spectral density defined as :

\begin{equation}
    S(f, t) = S_0 (f) \times g^2(f, t),
\end{equation}
where $S_0 (f)$ represents a stationary PSD, and $g(f, t)$ is an arbitrary function of time that introduces perturbations to the stationary PSD.

Using this approach, we extend the definition of the zero-mean GP presented in \autoref{eq:time_domain_stat_gp} by introducing time-varying coefficients $X_i(t)$ and $Y_i(t)$, where $X_i(t), Y_i(t) \sim \mathcal{N}(0, \sigma_i^2(t))$ and $\sigma_i^2 (t) = S_0(f_i) \times g^2 (f_i, t) \Delta f_i$. By applying the identity $\mathcal{N}(0, a^2 \sigma^2) = a\mathcal{N}(0, \sigma^2)$, the function $g(f, t)$ factors out of the Gaussian weights, yielding :

\begin{equation}
\begin{aligned}
    n(t) & = \sum_{i=1} ^N X_i (t_0) g(f_i, t)\sin(2\pi f_i t) + Y_i (t_0) g(f_i, t)\cos(2\pi f_i t)\\
    & = \sum_{i=1}^N \vec{\omega}_i \cdot \vec{\Phi}_i (t)
\end{aligned}
\end{equation}
with $\Phi_i (t) = g(f_i, t) \times \phi_i (t)$.

\begin{figure}[h]
	\centering
	\includegraphics[width=0.5\textwidth]{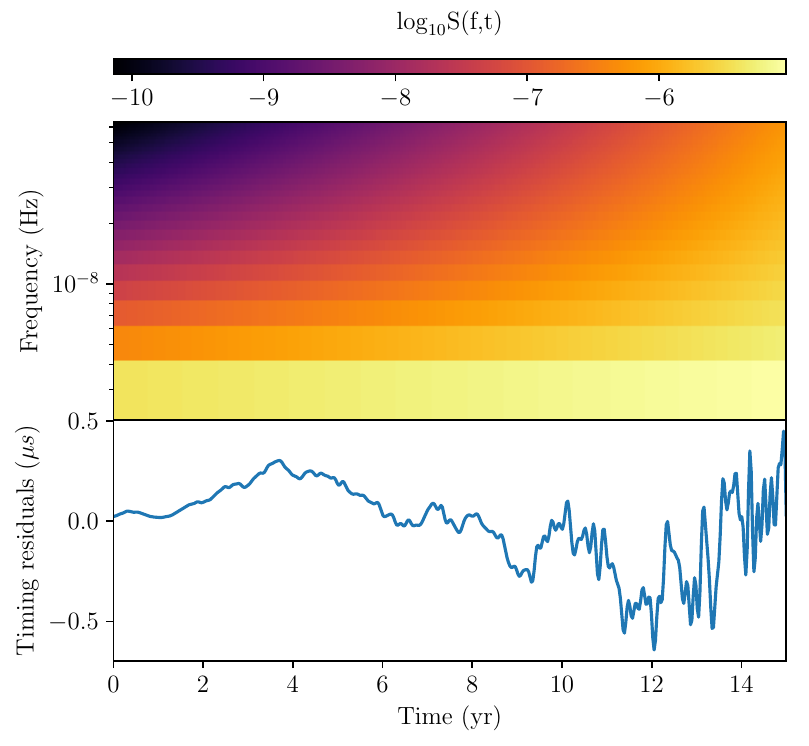}
	\caption{Example of a simulated non-stationary coloured noise for $\log_{10} A(t_0) = -14$ and $\gamma(t_0) = 2$ observed in the timing residuals of a PTA pulsar with 15 years of observation. The noise was simulated using a first order Taylor expanded power-law (see \autoref{eq:pl_ns_first_order}) with $a=0.8$ and $b=-1.2$ meaning that we have a noise with increasing amplitude and decreasing spectral index (i.e. the noise whitens with time). The top panel shows the evolutionary PSD $S(f, t)$ and the bottom panel shows the corresponding time domain representation.}
	\label{fig:pl_spectrum}
\end{figure}

This expression is identical to the stationary case except the basis $\Phi_i (t)$ can now be \mf{modulated} over time through the function $g(f_i, t)$. This \mf{modulation} of the basis gives a description of the non-stationarity of the signal (see \autoref{fig:pl_spectrum}). In this context, the relationship $\langle \vec{\omega}_i, \vec{\omega}_j \rangle = \langle \omega^2 \rangle _{ij}$ does not necessarily maintain a diagonal form because the weights are not expected to be uncorrelated. However, we proceed by making an assumption of diagonality. In other words, we assume that the basis  $\Phi_i (t)$ is the natural (eigen)basis of the process. In that case, the covariance matrix becomes:

\begin{equation}
\begin{aligned}
    C(t, t') & = \sum_{ij} \vec{\Phi}_i^\top (t) \langle \omega^2 \rangle _{ij} \vec{\Phi}_j(t') \\
    & \simeq \sum_i ^N \sigma_i ^2 g(f_i, t) g(f_i, t') \cos(2\pi f_i (t-t'))
\end{aligned}
\label{eq:nonstat_kernel}
\end{equation}

This kernel can be seen as a specific instance of non-stationary generalized spectral kernels (see \cite{samo2015generalized, remes2017nonstationary} and appendix). It is simplified by considering stationary frequency components $f_i$ accompanied by time-varying weights (a consequence of the assumption that $\langle \vec{\omega}_i, \vec{\omega}_j \rangle = \langle \omega^2 \rangle _{ij} = \sigma_i ^2 I_2 \delta_{ij}$ is diagonal). This assumption is reasonable if the evolution of frequencies $f_i$ of the basis $\Phi_i (t)$ is negligible during the considered time-span of observation $T$ (i.e. stationary phase). Such drifts in $f_i$ (typically chirp-like signals) would produce correlations between consecutive frequencies that are not accounted for in the expression of this kernel. Nevertheless, modeling the time dependent variance of each weight with this expression still captures information about non-stationary behaviors through the function $g(f, t)$.

The previous expression is valid for slowly evolving spectra that can be considered locally stationary. The notion of local stationarity is defined in \cite{mallat} as when the signal is approximately stationary over some sliding interval $\tau$. In our case, \mf{because we want to model slowly evolving spectra, the stationarity timescale $\tau$ should be comparable to the total time of observation $T$.} This condition translates to the function $g(f, t)$ as having most of its Fourier spectrum concentrated at frequencies around $1/T$. Essentially, we approximate non-stationary kernels by concentrating the 2D spectral measure $S(f, f')$ along the diagonal where $f = f'$ (see appendix \ref{sec:appendix_2dns}).

In the upcoming section, we propose conducting simulations to assess the impact of this approximation on parameter recovery. The primary goal is to estimate whether neglecting correlations between Gaussian weights in \autoref{eq:nonstat_kernel} significantly influences the inference of parameters within a Bayesian framework.

\subsection{Colored noise}

In PTA, pulsar noise is typically described with a power-law spectrum $S_0(f) \propto A f^{-\gamma}$, where $A$ represents the amplitude and $\gamma$ denotes the spectral index \cite{Agazie_noise_2023, Reardon_noise_2023, wm2}. Let us consider a time-dependent power spectral density with time-varying amplitude and spectral index :

\begin{equation}
S(f,t) = \frac{A^2(t)}{12 \pi^2} \bigg(\frac{f}{f_{yr}}\bigg) ^{-\gamma(t)} f_{yr}^{-3},
\end{equation}
\\
with $f_{yr}$ the frequency corresponding to one year. Any arbitrary functions of time for $A(t)$ and $\gamma (t)$ can be used as long as we ensure $S(f, t)>0$. We choose to express the amplitude and spectral index as:

\begin{equation}
\begin{aligned}
\log_{10} A(t) & = \log_{10} A(t_0) + \sum_k a_k P_k(t-t_0),\\
\gamma(t) & = \gamma(t_0) + \sum_k b_k P_k(t-t_0),\\
\end{aligned}
\end{equation}
where $t_0$ denotes the reference time, $a_k$ and $b_k$ are constant coefficients, and $P_k$ are polynomial functions (such as Chebyshev, spline, etc.). We can factor out the stationary PSD $S_0(f) \propto A^2 (t_0) f^{-\gamma(t_0)}$ from a time-dependent function $g(f, t)$:

\begin{equation}
\begin{aligned}
S(f,t) & = S(f, t_0) \times 10^{2\sum_k a_k P_k(t-t_0)} \bigg(\frac{f}{f_{yr}}\bigg)^{-\sum_k b_k P_k(t-t_0)}\\
& = S_0(f) \times g^2(f, t),
\end{aligned}
\end{equation}
providing an expression for the function $g(f, t)$ as presented in the previous section. According to \autoref{eq:nonstat_kernel}, we then have:

\begin{equation}
\begin{cases}
\begin{aligned}
C(t, t') & = \sum_i \sigma_i^2 g(f_i, t) g(f_i, t') \cos(2\pi f_i (t-t'))\\
\sigma_i^2 & \propto A^2(t_0) f_i^{-\gamma(t_0)} \Delta f_i\\
g(f, t) & = 10^{\sum_k a_k P_k(t-t_0)} \bigg(\frac{f}{f_{yr}}\bigg)^{-\sum_k \frac{b_k}{2} P_k(t-t_0)}
\end{aligned}
\end{cases}
\end{equation}

In the next sections of this paper, we will only utilize the first-order term of this expression. We will assume slowly varying properties of the signal.

\section{Injection recovery test}

\subsection{Likelihood}
\label{sec:inference}

To perform inference of the parameters, we need to define a likelihood. The GP model previously described assumes that the noise in the data is Gaussian. Hence, for a time series $s(\vec{t})$ of length $N$ measured at times $\vec{t}$, we will consider a Gaussian likelihood of the form :

\begin{equation}
\mathcal{L}(s|\vec{\theta}) = \frac{1}{|2\pi C(\vec{\theta})|^{\frac{1}{2}}} \exp \bigg \{ -\frac{1}{2} \sum_i \sum_j s(t_i) C^{-1}(\vec{\theta}, t_i, t_j) s(t_j) \bigg \}
\end{equation}
where $C^{-1}$ is the inverse of the covariance matrix $C$, $\vec{\theta}$ the vector of the model parameters and $|.|$ denotes the determinant.

In the Bayesian framework, we define prior probability distributions $\pi(\vec{\theta})$ for the parameters $\vec{\theta}$ of the model. Combining the prior distributions with the likelihood, we can update our knowledge about $\vec{\theta}$, defining the posterior distribution $p(\vec{\theta}|s)$ :

\begin{equation}
p(\vec{\theta}|s) = \mathcal{L}(s|\vec{\theta}) \pi(\vec{\theta}) / \mathcal{Z},
\end{equation}
with $\mathcal{Z}$ the so-called \textit{evidence} of the model acting as a normalizing constant.

The key quantity to evaluate the significance of model $A$ against model $B$ in Bayesian analysis is the Bayes factor $\mathcal{B}^A_B$. It is defined as the ratio of the evidences

\begin{equation}
    \mathcal{B}^A_B =\frac{\mathcal{Z}_A}{\mathcal{Z}_B}.
\end{equation}

In this paper, the posterior distribution is sampled using Markov Chain Monte Carlo (MCMC) sampler \texttt{PTMCMC} \cite{justin_ellis_2017_1037579} or nested sampling library \texttt{dynesty} \cite{skilling}.

\subsection{Time-varying filter}
\label{sec:gen_nonstat_rn}

\mf{In order to test our model, we want to simulate a colored noise with evolving properties in time using a method that is independent of the one developed in the previous section. Various methods are available to generate such noise, for example, one can utilize auto-regressive models \cite{cornish_wdm}.} Here we chose to generate non-stationary data with a time-varying filter that allows one to control the amplitude and spectral index of the noise at each instant \cite{mallat, priestley}.

\mf{We first generate an evenly sampled white noise series $y \sim \mathcal{N}(0, 1)$ of size $N_s+1$ with a cadence $\Delta t$ and a total duration $N_s \Delta t$. Then, the time series $y$ is filtered using a time-dependent filter $\hat{h}(f, t)$.} The filtering is performed in the frequency domain. For a colored noise, we define a filter in Fourier domain $\hat{h}(f, t)$ at time $t$ as:

\begin{equation}
\hat{h}(f, t) = 
\left\{\begin{matrix}
\begin{aligned}
& \ \ \ \ \ \ \ \ \ \ \ \ \ 0\ \ \ \ \ \ \ \ \ \ \ \ \ \ \ \ \textrm{for}\ f<1/N_s \Delta t \\ 
& g(f, t) H_0 \bigg(\frac{f}{f_{yr}}\bigg)^{-\alpha_0/2}\ \ \textrm{for}\ f \geq 1/N_s \Delta t
\end{aligned}
\end{matrix}\right.
\end{equation}
where $H_0$ is the gain of the filter, and $\alpha_0$ is the slope. The gain and slope can be controlled over time using the function $g(f, t)$.

For a time series $y$ and its Fourier transform $\hat{y}(f)$, the output $y'(t)$ of the filter is :

\begin{equation}
	y'(t) = \mathcal{F}^{-1} (\hat{h}(f, t) \hat{y}(f)),
\end{equation}

with $\mathcal{F}^{-1}$ the inverse Fourier transform.

Finally, we fit a quadratic polynomial of the form $a + bt + ct^2$ to the generated time series. \mf{The obtained polynomial is then subtracted from the time series. This final step mimics the timing model fit performed in PTA data \cite{wm1, wm2}, which accommodates the quadratic spin-down (deceleration) of the pulsar rotation while also eliminating part of the moving average component attributed to the non-stationarity.}

\subsection{Setting the model and priors}

To test the model, we aim to estimate the Bayes factor for the non-stationary case versus the stationary case given the generated data. As mentioned earlier, we want to restrict ourselves to first-order variations in time of the power-law spectrum amplitude and spectral index. \mf{We evaluate the covariance matrix in \autoref{eq:nonstat_kernel} on a discrete set of frequencies $f_i = i/\tau$ with $i=[1, ..., N_f]$ an integer, $N_f$ the size of the basis and $\tau$ the considered stationarity time-scale}. In this context, the function $g(f_i, t)$ can be expressed as :

\begin{equation}
    \log _{10} g(f_i, t) = a \frac{(t-t_0)}{T} -  \frac{1}{2} b \frac{(t-t_0)}{T} \log _{10} (f_i/f_{yr}).
\label{eq:nonstat_red_noise}
\end{equation}
\mf{where $T$ is the total time of observation, $f_{yr}=1/(1yr)$ is a reference frequency and $t_0$ is the initial time.}

\begin{table}[b]
\caption{\label{tab:priors}
Prior probability distributions used for the parameters of a non stationary coloured noise.}
\begin{tabular}{c|c}
Parameter & Prior probability\\
\hline
$\log_{10} A(t_0)$ & Uniform(-18, -10) \\
$\gamma (t_0)$ & Uniform(1, 6)\\
$a$ & $\mathcal{N}(0, 1)$ \\
$b$ & $\mathcal{N}(0, 1)$
\end{tabular}
\end{table}

In this work, we chose to fix $t_0$ at the start time of the entire dataset to directly measure the evolution of amplitude and spectral index from starting values $\log_{10} A(t_0)$ and $\gamma(t_0)$.

To perform inference of the parameters, we need to set prior probability distributions. For a slowly evolving signal observed within a time interval $T$, we assume that \mf{the stationarity time-scale $\tau \simeq T$ so the chosen function $g(f_i, t)$ must have most of its Fourier spectrum concentrated around the lowest frequency $1/T$.} This ensures that the 2D spectral measure is concentrated on the diagonal $f=f'$, minimizing leakage between different frequencies (see \autoref{sec:appendix_2dns}). If not, $g(f_i, t)$ is rapidly evolving, and the assumption of local stationarity is no longer valid. \mf{}

\begin{figure}[h]
	\centering
	\includegraphics[width=0.5\textwidth]{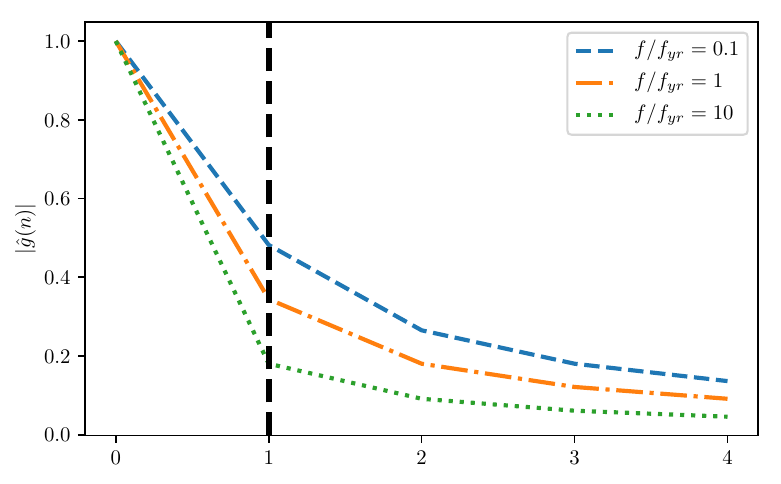}
	\caption{Norm of the Fourier transform of $g(f_i, t)$ expressed in \autoref{eq:nonstat_red_noise} for $a=1$, $b=1$ and three ratios $f_i/f_{yr}$. The x-axis corresponds to the harmonics of $1/T$. We show $|\hat{g}(n)| / |\hat{g}(0)|$ for better visibility and comparison.}
	\label{fig:g_fourier}
\end{figure}

\mf{In \autoref{fig:g_fourier}, we display the Fourier spectrum of $g(f_i,t)$ on the interval $[t_0,t_0+T]$, denoted as $\hat{g}(n)$ where $n$ corresponds to the harmonics of $1/T$. We can observe that for parameter values $a = b = 1$, the condition is roughly fulfilled. However, as we decrease the ratio $f_i/f_{yr}$ in the expression of $g(f_i, t)$, the condition is less respected. This implies that the Fourier transform $\hat{g}(n)$ of the function $g(f_i,t)$ has a wider spread at lower $f_i$, leading to more significant changes occurring within timescales shorter than $T$. Thus, we choose $a=b=1$ to represent typical parameter values where the assumption of a slowly evolving spectrum begins to break down for $T\sim10yr$. We decide to set a normal distributed prior $\mathcal{N}(0, 1)$ for $a$ and $b$ (see \autoref{tab:priors}) to confine their values around the interval $[-1,1]$}. The inference of parameters may not be precise enough to accurately estimate the evolutionary PSD of the process within the Bayesian scheme (resulting in wide posterior uncertainty). In this sense, this model remains an approximation. Nevertheless, we should be able to capture the non-stationarity of the process through the evaluation of the Bayes factor if $a$ and $b$ deviate too much from 0. This last point is crucial to test whether we are actually dealing with non-stationary noise.

\subsection{Simulation results}
\label{sec:sim_results}

Following the procedure presented in \autoref{sec:gen_nonstat_rn}, we generate 1000 simulations with 500 samples and 20 years of data containing non-stationary red noise with a low white noise level of $10^{-9}s$. For each simulation, the parameters $a$ and $b$ are drawn from their respective prior probability distribution, while the initial gain $H_0$ and slope $\alpha_0$ are fixed at $10^{-7}$ and $3$, respectively. Inference is performed using the expressions of the likelihood and posterior probability presented in \autoref{sec:inference}. The prior distributions we use are given in \autoref{tab:priors}. We employ nested sampling to explore the parameter space for both stationary and non-stationary models to estimate the Bayes factors \cite{skilling}. We denote the Bayes factor for non-stationarity against stationarity of noise as $\mathcal{B}^{NS}_S$.

\begin{figure}[h]
	\centering
	\includegraphics[width=0.5\textwidth]{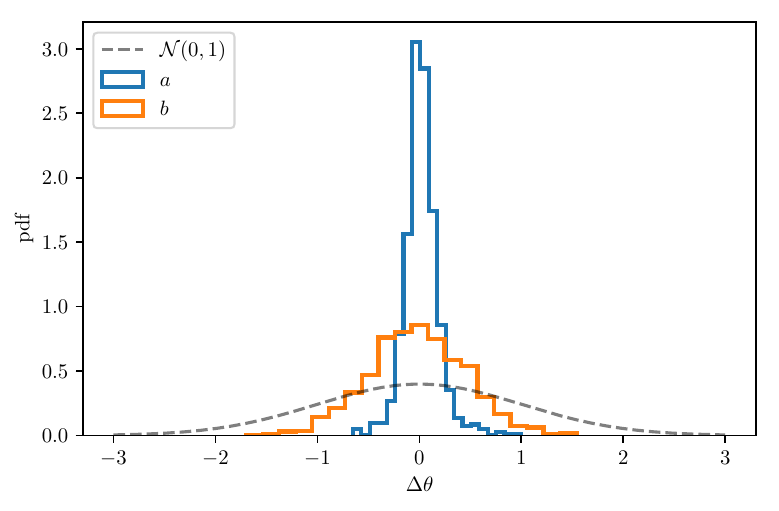}
	\caption{Normalized histogram of the differences $\Delta \theta$ between injected and recovered maximum likelihood parameter value.}
	\label{fig:param_errors}
\end{figure}

 \autoref{fig:param_errors} illustrates the precision of non-stationary parameter recovery. We present a histogram showing the differences in each simulation between the injected and recovered maximum likelihood values of parameters $a$ and $b$. The uncertainties are narrower than the prior distribution, indicating that the model is informative. \mf{However, the histogram for $b$ appears broader than parameter $a$, which makes an accurate estimate of the evolution of the spectral index challenging.}

\begin{figure}[h]
	\centering
	\includegraphics[width=0.5\textwidth]{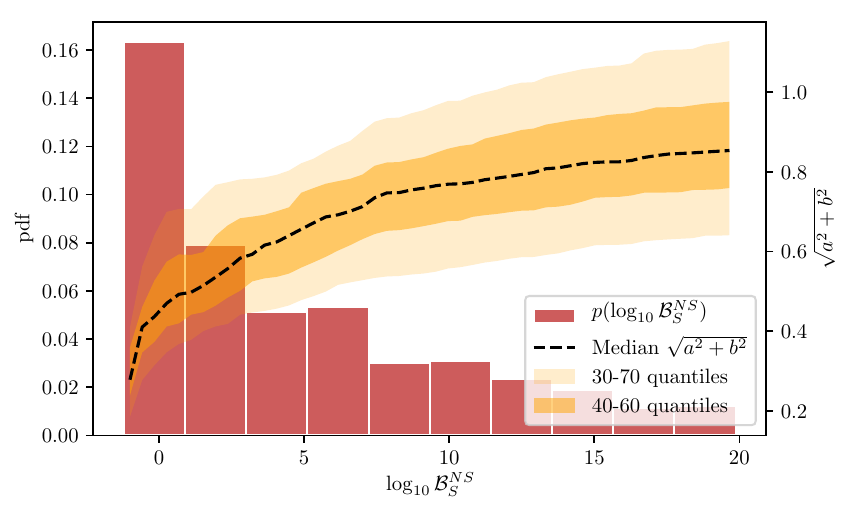}
	\caption{Normalized histogram of the Bayes factor $\mathcal{B}^{NS}_S$ non-stationary vs stationary noise obtained from each simulation. We define the level of non-stationarity $\sqrt{a^2 + b^2}$ combining parameters $a$ and $b$. We plot the distribution obtained from their recovered maximum likelihood values for the corresponding simulation and Bayes factor in the x-axis.}
	\label{fig:log10_bfs}
\end{figure}

\autoref{fig:log10_bfs} demonstrates the ability to accurately predict non-stationarity through the evaluation of the Bayes factor $\mathcal{B}^{NS}_S$. We present a histogram of the Bayes factors evaluated for each simulation. The distribution peaks at low Bayes factors, which corresponds to the fact that we draw $a$ and $b$ from $\mathcal{N}(0, 1)$, hence they are often near zero. If $a$ and $b$ are small, the stationary model should be equally preferred. Additionally, we observe that the Bayes factor rapidly increases if $a$ and $b$ significantly deviate from zero. \mf{However, in this simulation, due to the large uncertainties on the recovery of $b$, the information it provides is limited, and the recovered maximum likelihood value does not effectively convey this lack of information (see \autoref{sec:appendix_sim}). In \autoref{fig:log10_bfs}, the non-stationarity level $\sqrt{a^2 + b^2}$ and the increase in the Bayes factor are predominantly influenced by the value of parameter $a$.}

\section{Results}

The results section will focus on the search for a GWB in the second data release of the EPTA collaboration \cite{wm1}. The following analyses use the same methods, priors, and noise models presented in \cite{wm2, wm3}. We use the 24.7 years of dataset referred to as \texttt{DR2full} in \cite{wm3}. The search is for a GWB with a power-law spectrum modeled as stochastic noise spatially correlated between pairs of pulsars, following the so-called Hellings-Downs correlation pattern \cite{Allen_2023}. The individual pulsar noise models are constructed from a combination of RN and DM noise, also modeled with power-law spectra \cite{wm2}. These pulsar noise models are built using \texttt{ENTERPRISE} \cite{enterprise}.

First, we study the GWB using the stationary formulation of the covariance matrix, as described in \cite{wm3}, but across different time slices of the dataset. Then, we investigate the GWB using the non-stationary formulation of the covariance that was developed in the previous sections

\subsection{Time-slice analysis of GWB}

\begin{figure}[ht]
\begin{tikzpicture}
\draw[-{latex}, black, ultra thick] (-2.5,1) -- (6.2,1);
\draw[dashed, black] (-2.2,0.2) -- (-2.2,1.8);
\filldraw[blue, very thick, opacity=0.5] (-2.2,0.85) rectangle (1.8,1.25);
\filldraw[orange, very thick, opacity=0.5] (-0.2,0.75) rectangle (3.8,1.15);
\filldraw[green, very thick, opacity=0.5] (1.8,0.65) rectangle (5.8,1.05);
\draw [thick, decorate,decoration={brace,amplitude=10pt}]
(5.8,0.5) -- (-2.2,0.5);
\draw (1.8,-0.3) node {$T$};
\draw (-2.5,0.6) node {$t_0$};
\draw [thick, decorate,decoration={brace,amplitude=10pt}]
(-2.2,1.5) -- (1.8,1.5);
\draw (-0.2,2.3) node {$T/2$};
\end{tikzpicture}
\caption{Timeline representation (to scale) of the three analyzed slices represented by the colors blue, orange and green with respect to the full time of observation $T$.}
\label{fig:timeline}
\end{figure}
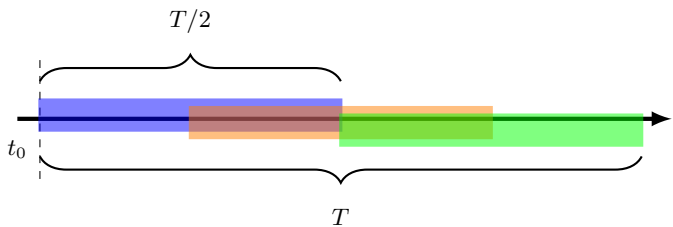

Time-slice analyses of the GWB were previously conducted by the Parkes Pulsar Timing Array collaboration for their 3rd data release \cite{ppta_dr3}, where the dataset was segmented into several slices and analyzed as a stationary GWB in each slice. In this section, we perform a similar analysis on three different slices of our dataset. For $T$, the total time span of observation, and $t_0$, the initial time, we analyze the slices [$t_0$, $T/2$], [$T/4$, $3T/4$], [$T/2$, $T$], as illustrated in \autoref{fig:timeline}.

\begin{figure}[h]
	\centering
	\includegraphics[width=0.5\textwidth]{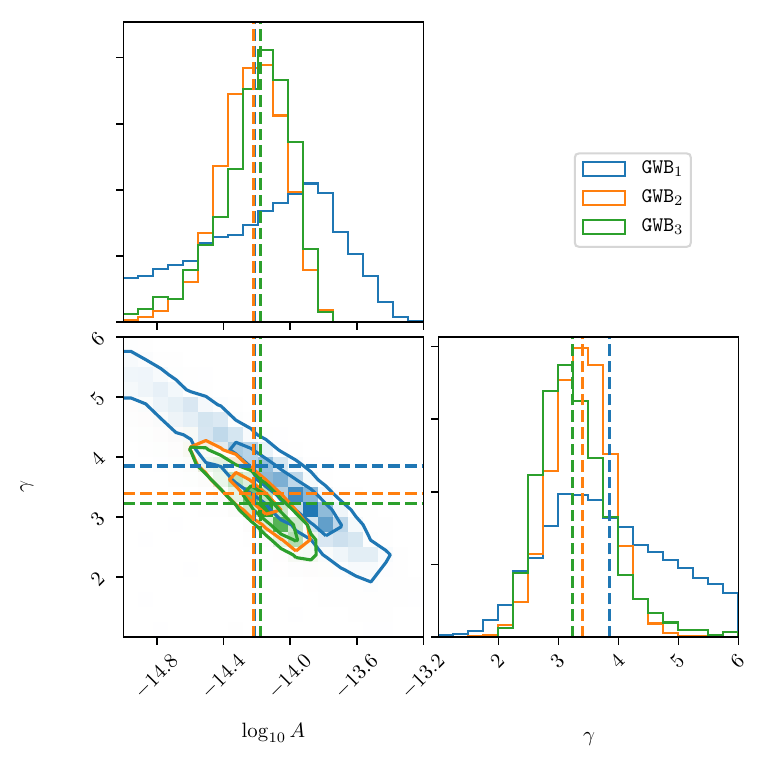}
	\caption{2D posterior distributions of GWB parameters analyzed in three time slices. The contours on the 2d histograms are the $68\%$ and $95\%$ credible regions.}
	\label{fig:gwb_corner_slice}
\end{figure}

In all three slices, we search for a stationary GWB with a power-law spectrum on a 30-frequency linearly spaced Fourier basis with the lowest frequency $2/T$, allowing the spectral index $\gamma$ and amplitude $\log_{10}A$ to vary (see \cite{wm3} for details). We denote the recovered GWB posteriors in slices 1, 2, and 3 as \texttt{GWB}$_1$, \texttt{GWB}$_2$, and \texttt{GWB}$_3$ respectively. The results are presented in \autoref{fig:gwb_corner_slice}.

The \texttt{GWB}$_1$ posterior appears to be much broader than the others. This is expected because the cadence of early observations was lower than those of today \mf{(some pulsars may have a cadence of 1 point per month in the early data versus 1 every 3 days in the latest). The level of white noise is directly influenced by the cadence of observation\footnote{For an evenly sampled dataset with measurement uncertainties $\sigma$ and red noise, the PSD can be expressed as $S(f) = A f^{-\gamma} + 2\sigma^2 \Delta t$ where $\Delta t$ is the cadence of observation that directly influences the tail of the spectrum \citep{Moore_2015}.}. The interplay between the powerlaw spectrum of red noise and the flat white noise tail may strongly influence the estimate of the red noise parameters}. Moreover, the quality of the backends processing raw data from the radio observatories has improved over time. Their ability to simultaneously observe incoming photons at multiple frequency bands (larger bandwidth) allows for a better characterization of the DM noise due to pulses traveling through the interstellar medium \cite{wm2}. The latter induces a delay in the time of arrival of the pulses with an amplitude that is inversely proportional to the square of the incoming photon's frequency. Its characterization requires good multiband observing capabilities; otherwise, it could be mistaken for RN and contaminate GWB parameter estimation. Improved cadence and noise characterization should provide better constraints on the GWB posteriors, which could explain why \texttt{GWB}$_1$ appears less informative than \texttt{GWB}$_2$ and \texttt{GWB}$_3$. Moreover, if some of the individual pulsar noises are non-stationary, they may leak into the GWB signal. This last point will be explored in the next subsection. It is important to note that this first slice, \texttt{GWB}$_1$, corresponds to the first data release from EPTA, where no GWB was less significant. The recovered posteriors are similar to what was observed in \cite{epta_dr1}.

\begin{figure}[h]
	\centering
	\includegraphics[width=0.5\textwidth]{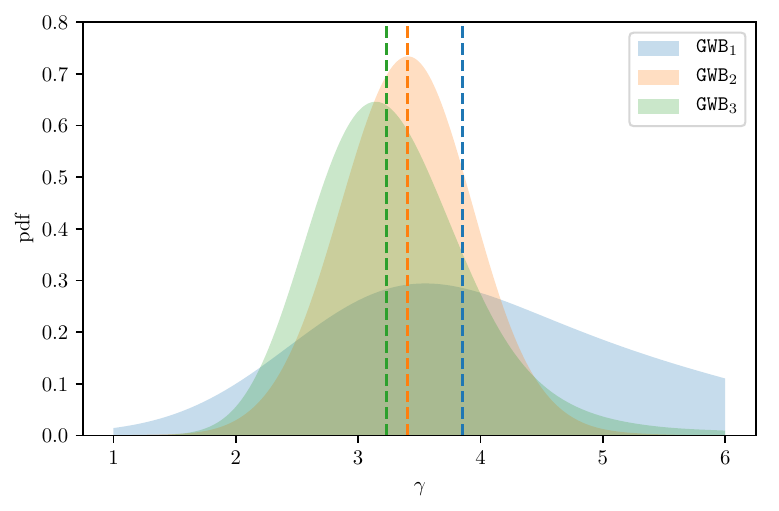}
	\caption{1D marginalized posterior distribution of GWB spectral index $\gamma$ analyzed in three time slices.}
	\label{fig:gamma_slice}
\end{figure}

The \texttt{GWB}$_2$ and \texttt{GWB}$3$ posteriors nearly overlap and closely resemble the recovered posteriors using the latest 10 years of data in \cite{wm3}. We observe a slight shift of the spectral index $\gamma$ towards lower values. \autoref{fig:gamma_slice} displays the 1D posteriors of $\gamma$ for the three slices, indicating an estimated decrease in $\gamma$ of approximately 0.66 from the median values. Regarding the amplitude, the median values show little to no evolution. However, evaluating the evolution of $\gamma$ and $\log_{10}A$, especially for \texttt{GWB}$_1$, is challenging due to the large uncertainties.

One caveat of this analysis is that we lose information about the low-frequency range after performing the slices. Indeed, the lowest sampled frequency bin is at frequency $1/T$, and increasing $T$ raises its value. In PTA, the lowest frequencies contain important information when searching for a GWB. Additionally, this analysis prevents us from estimating a Bayes Factor in favor of non-stationary behavior against stationarity. The model presented in this paper should address both issues.

\subsection{Non-stationary powerlaw spectrum GWB}

We will now analyze the data using a GWB model with HD correlations and an evolutionary power-law spectrum. In \autoref{subsec:ns_gp}, we developed an expression for such a model.

The previous time-slice analysis only reveals small variations in the amplitude and spectral index of the power-law. For this reason, we chose to limit ourselves to the simplest possible model where the evolution of amplitude and spectral index is approximated to first order in time. We have:

\begin{equation}
\begin{aligned}
\log_{10} A(t) & \sim \log_{10} A(t_0) + a(t-t_0) / T,\\
\gamma(t) & \sim \gamma(t_0) + b (t-t_0) / T,\\
\end{aligned}
\label{eq:pl_ns_first_order}
\end{equation}
where $t_0$ is the minimum time of the dataset, $a$ the parameter controlling the evolution of log-amplitude and $b$ the parameter controlling the evolution in $\gamma$.

\begin{figure}[h]
	\centering
	\includegraphics[width=0.5\textwidth]{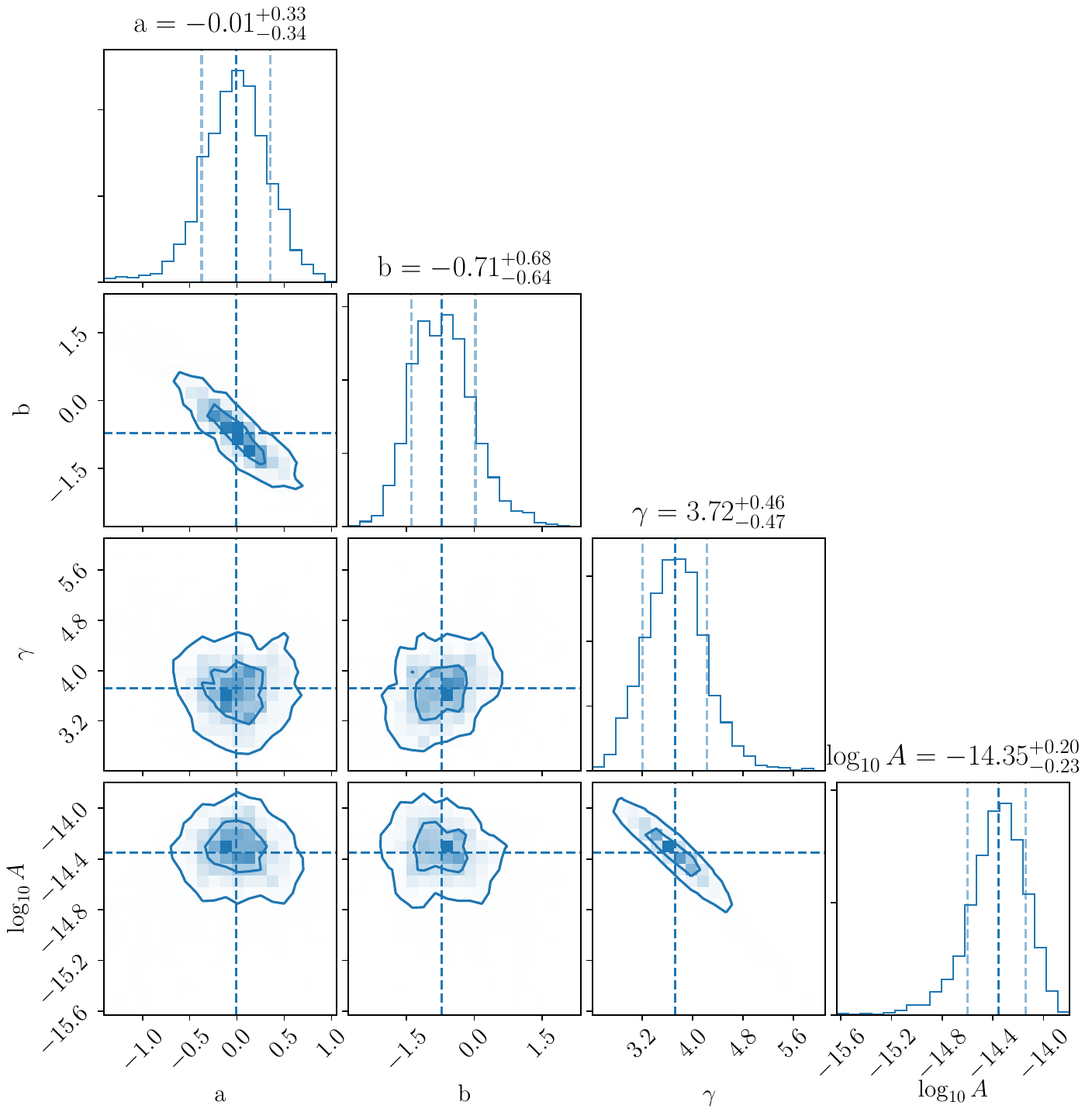}
	\caption{Corner plot of non-stationary GWB parameters inferred from the EPTA dataset. The contours on the 2d histograms are the $68\%$ and $95\%$ credible regions. The dashed lines on the 1d histograms show the $14/50/86\%$ percentiles.}
	\label{fig:gwb_nonstat}
\end{figure}

We performed the analysis using a normal prior $\mathcal{N}(0, 1)$ on $a$ and $b$ and show the recovered posteriors in \autoref{fig:gwb_nonstat}. We find a GWB with $\log_{10} A(t_0) = -14.4_{-0.2}^{+0.2}$, $\gamma (t_0)=3.7_{-0.5}^{+0.5}$, $a=0.0_{-0.3}^{+0.3}$, and $b=-0.7_{-0.6}^{+0.7}$, fitting a spectrum with a constant amplitude and decreasing spectral index. This behavior aligns with the estimated evolution of the parameters using the medians of the 1D posteriors in the slice analysis. \mf{Still, the uncertainties on the estimate of parameter $b$ are very large and $b=0$ lies within the $1-\sigma$ credible interval.} For this GWB model, we estimate a Bayes factor of non-stationarity against stationarity $\mathcal{B}^{NS}_{S} = 1.5$, indicating no evidence for a non-stationary GWB. \mf{The apparent evolution of the spectral properties of this process is likely due to improvements in receiver-backend systems and the enhanced cadence of observation over the years. As demonstrated in the previous section, the inferred posteriors of the GWB parameters in the oldest slice of our dataset are very uninformative. This limitation may impede our ability to characterize the time-dependent features of the spectrum effectively.}

\subsection{Non-stationary individual noise}

Earlier we mentioned the possible non-stationarity of individual pulsar noise. In the introduction, we presented the two dominating sources of time-correlated noise in pulsar timing :the RN, due to stochastic variation of the pulsar's spin rate, and the DM noise, due to the variation of the interstellar medium electron density throughout the travel of the pulses to Earth \cite{wm2}. \mf{Another source of frequency dependent noise is identified}, chromatic noise, with an amplitude proportional to $\nu^{-4}$ with $\nu$ the incoming photon frequency. This type of noise is assumed to be stationary in PTA data analysis. Incorrect modeling of pulsar noise could deteriorate the inference of the GWB parameters. 

In this section, we compare the stationary formulation of the custom noise models presented in \cite{wm2} to the non-stationary formula of section \autoref{subsec:ns_gp} in order to assess the stationarity of pulsar noises. Every noise spectrum is modeled as a power-law spectrum. We only test the first order model with function $g(f, t)$ given by \autoref{eq:nonstat_red_noise} to allow each noise component to vary with time. We perform MCMC analysis using product-space sampling \cite{hypermodel} to estimate the Bayes factor $\mathcal{B}^{NS}_{S}$. Results are gathered in \autoref{tab:psr_noise_bf}.

According to the distribution of Bayes factors, it appears that the pulsar noise do not exhibit first order variations in time, with the exception of J1713+0747 for which $\log_{10}\mathcal{B}^{NS}_{S}=4.3$, hence favoring the non-stationary model.

Given the noise parameter values in \autoref{tab:1713_noise}, it seems that the recovered non-stationarity originates from the DM noise with a relatively well constrained $a_{DM}-0.7 _{-0.2}^{+0.2}$ indicating that the DM amplitude decreases with time and $b_{DM}=0.4 _{-0.6}^{+0.6}$ indicating an increasing spectral index \mf{(though with large posterior uncertainties). For the RN, the $a_{RN}$ and $b_{RN}$ parameters are consistent with zero within the $1-\sigma$ credible interval suggesting stationary behavior.}

\mf{However, J1713+0747 is known to have challenging noise properties \cite{NG_noise_budget, Lam_2021, singha} and non-stationary DM events were already reported for this pulsar in \cite{lentati_2016}. In \cite{wm2}, two exponential dip events were reported in the timing residuals of J1713+0747. An exponential dip is a deterministic DM-like effect that could be induced by extreme scattering events in the interstellar medium or a sudden change in the pulse profile of the pulsar, introducing transitory frequency dependent timing delays that can last up to a few months \cite{Shannon_2016, Coles_2015}. In \autoref{tab:1713_noise}, we see that the inclusion of the exponential dips in our model removes the apparent non-stationary behavior of J1713+0747, now showing $a_{DM}=-0.2 _{-0.2}^{+0.3}$ and $b_{DM}=-0.4 _{-0.7}^{+0.7}$. The Bayes factor for non-stationary noise versus stationary noise after the inclusion of exponential dip events is now $\log_{10}\mathcal{B}^{NS}_{S}=-0.6$.}

\mf{Although not the best description, the non-stationary model was able to hint towards the presence of time-dependent features in the timing residuals of pulsar J1713+0747 through the evaluation of the Bayes factor. We tried searching again for a GWB in the EPTA dataset, now considering the non-stationarity (exponential dips) of J1713+0747, and found no difference in the estimate of the Bayes factor in favor of HD or in the inferred GWB parameters with respect to \cite{wm3}. The stationary model of individual pulsar noise seems to be enough an approximation when searching for a GWB.}

\begin{table}
\begin{tabular}{c|c}
Pulsar & $\log_{10}\mathcal{B}^{NS}_{S}$\\
\hline
J0030+0451 & 0.0 \\
J0613-0200 & -0.6 \\
J0751+1807 & -0.3 \\
J0900-3144 & -0.3 \\
J1012+5307 & -1.2 \\
J1022+1001 & -0.2 \\
J1024-0719 & -0.3 \\
J1455-3330 & 0.0 \\
J1600-3053 & 0.0 \\
J1640+2224 & -0.2 \\
\textbf{J1713+0747} & \textbf{4.3}/-0.6\\
J1730-2304 & 1.1 \\
J1738+0333 & 0.1 \\
J1744-1134 & -0.5 \\
J1751-2857 & -0.2 \\
J1801-1417 & 0.3 \\
J1804-2717 & 0.2 \\
J1843-1113 & -0.4 \\
J1857+0943 & -0.4 \\
J1909-3744 & 0.7 \\
J1910+1256 & 0.0 \\
J1911+1347 & -0.4 \\
J1918-0642 & -0.3 \\
J2124-3358 & -0.3 \\
J2322+2057 & 0.0 \\
\end{tabular}
\caption{Bayes factors for non-stationary against stationary noise models. The pulsar with $\log_{10}\mathcal{B}^{NS}_{S}>2$ is shown in bold font. We observe that, in general, pulsars do not exhibit evidence of non-stationary noise, except for J1713+0747, for which we show the two values of $\log_{10}\mathcal{B}^{NS}_{S}$ obtained with (left) and without (right) including the exponential dip events in the model. The case of J1713+0747 is discussed in the text.}
\label{tab:psr_noise_bf}
\end{table}

\begin{table}
\begin{tabular}{c|c|c|c}
Parameter & Median (NS) & Median (NSed)& Median (Sed) \\
\hline
\hline
 & & \\
$\gamma_{DM}$ & $1.0 _{-0.4}^{+0.4}$ & $1.9 _{-0.6}^{+0.6}$ & $1.6 _{-0.2} ^{+0.2}$\\[0.1cm]
$\log_{10}A_{DM}$ & $-13.0 _{-0.1}^{+0.1}$ & $-13.6 _{-0.2}^{+0.2}$ & $-13.7 _{-0.04} ^{+0.04}$\\[0.1cm]
$a_{DM}$ & $-0.7 _{-0.2}^{+0.2}$ & $-0.2 _{-0.2}^{+0.3}$ & - \\[0.1cm]
$b_{DM}$ & $0.4 _{-0.6}^{+0.6}$ & $-0.4 _{-0.7}^{+0.7}$ & - \\[0.1cm]
\hline
 & & \\
$\gamma_{RN}$ & $3.2 _{-0.8}^{+0.8}$ & $ 3.2 _{-0.8}^{+0.8}$ & $3.5 _{-0.6} ^{+0.6}$\\[0.1cm]
$\log_{10}A_{RN}$ & $-13.9 _{-0.5}^{+0.4}$ & $-14.0 _{-0.5}^{+0.4}$ & $-14.3 _{-0.3} ^{+0.3}$ \\[0.1cm]
$a_{RN}$ & $-0.3 _{-0.5}^{+0.5}$ & $-0.2 _{-0.5}^{+0.5}$ & - \\[0.1cm]
$b_{RN}$ & $-0.2 _{-0.9}^{+0.9}$ & $-0.3 _{-0.9}^{+0.9}$ & - \\[0.1cm]
\end{tabular}
\caption{Table of median noise parameters (RN and DM) inferred from pulsar J1713+0747 timing data for the non-stationary model (NS), non-stationary model with exponential dips (NSed) and the stationary model with exponential dips (Sed). The displayed quantiles are 0.14 and 0.86.}
\label{tab:1713_noise}
\end{table}

\subsection{Eccentric single source as non-stationary GWB}

If a binary system of SMBH is eccentric, it will radiate polychromatic GWs, contrary to the monochromatic circular binaries \cite{peters_mathews_ecc}. On average, the GW signal can be decomposed as a sum of frequencies, which are the harmonics of the orbital frequency of the binary system. The waveform of an eccentric GW shows that the instantaneous frequency of the signal changes significantly within one orbital period \cite{Susobhanan_2020, Taylor_ecc_2016}.

\begin{figure}[h]
	\centering
	\includegraphics[width=0.5\textwidth]{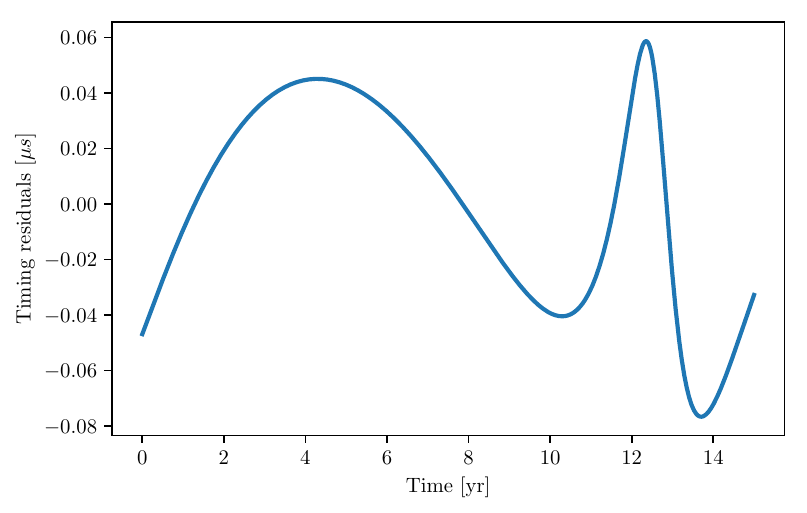}
	\caption{One period of an eccentric signal (Earth term only) with eccentricity 0.6 and orbital period of 15 years We see that the instantaneous frequency significantly changes within one period. The waveform was generated using \texttt{ENTERPRISE} extensions \cite{enterprise_extensions}.}
	\label{fig:ecc_resid}
\end{figure}

In \cite{wm4}, it was shown that single binaries could be mistaken for a GWB with a power-law spectrum. For eccentric binaries with orbital periods comparable to the total time of observation of a PTA, the instantaneous change in frequency of the GW signal can be seen as a GWB with evolving spectral index $\gamma(t)$ (see \autoref{fig:ecc_resid}). Indeed, the $\gamma$ controls the balance between low and high frequencies of the power-law spectrum and the presence of an eccentric signal would introduce time-dependent features. GW sources of higher eccentricity and higher amplitude would introduce more non-stationarity because of their stronger frequency evolution or higher signal to noise ratio (SNR). In this section, we want to investigate what we would see when a stationary GWB is influenced by an eccentric single source with low SNR.

\begin{figure}[h]
	\centering
 	\includegraphics[width=0.5\textwidth]{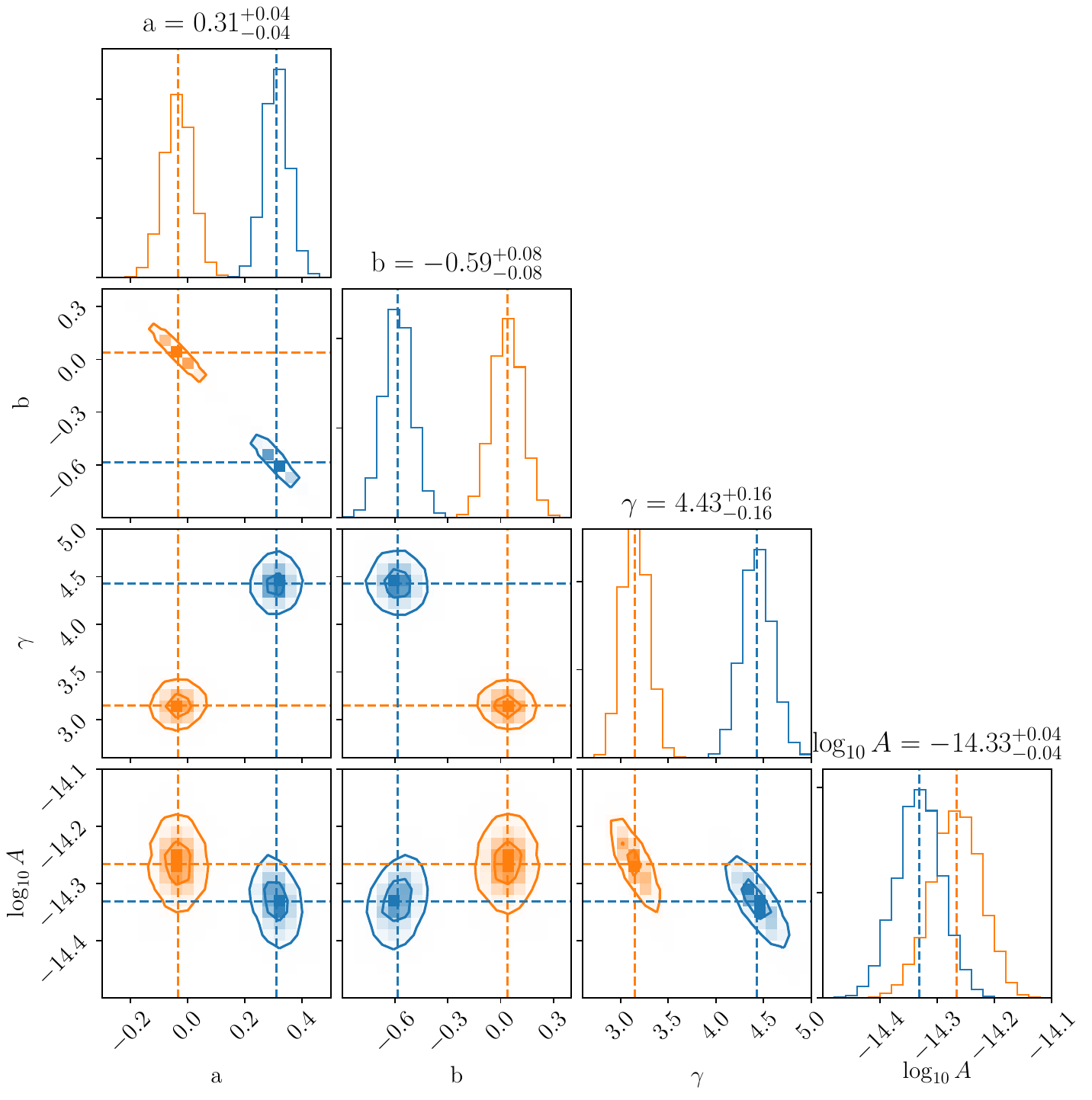}
	\caption{Corner plot of non-stationary GWB parameters. In orange, we show the results for a simulated stationary GWB with $\log_{10}A=-14.2$ and $\gamma = 3$. In blue, we show the results for the same realisation of GWB "contaminated" by an eccentric single source with 15 years orbital period and eccentricity $e_0=0.5$. The dashed line are the medians. The displayed values of parameters correspond to the medians of the orange histograms. The contours on the 2d histograms are the $68\%$ and $95\%$ credible regions.}
	\label{fig:gwb_ecc}
\end{figure}

We create a simulation with 10 pulsars (isotropically distributed in the sky), 15 years of observation and only white noise at a level of $10^{-7}s$. The simulation contains a stationary GWB with $\log_{10} A = -14.2$ and $\gamma = 3$. We generate two cases: (i) only a stationary GWB is present in the data (ii) the same stationary GWB is accompanied by an eccentric source located at the Virgo galaxy cluster with amplitude $h=10^{-13.5}$ and orbital period of 15 years (here corresponding to SNR $\simeq 1.8$). \mf{We produce 8 datasets with different eccentricities of the source $e$ ranging from 0.1 to 0.8}. We analyzed these datasets with the non stationary model presented in \ref{subsec:ns_gp}, considering HD correlations. In \autoref{fig:gwb_ecc}, we show that the presence of an eccentric source \mf{with $e=0.5$} introduces non stationarity in the inferred GWB parameters. For case (i), parameters $a$ and $b$ are near zero meaning that we are dealing with a stationary signal. In case (ii), where the eccentric signal is introduced, they deviate significantly from zero. Note that in our simulation, the parameter $b$ is negative (i.e. $\gamma$ is decreasing) showing that some specific realizations of eccentric binaries can somewhat reproduce the behavior observed in the real data.

\begin{figure}[h]
	\centering
 	\includegraphics[width=0.5\textwidth]{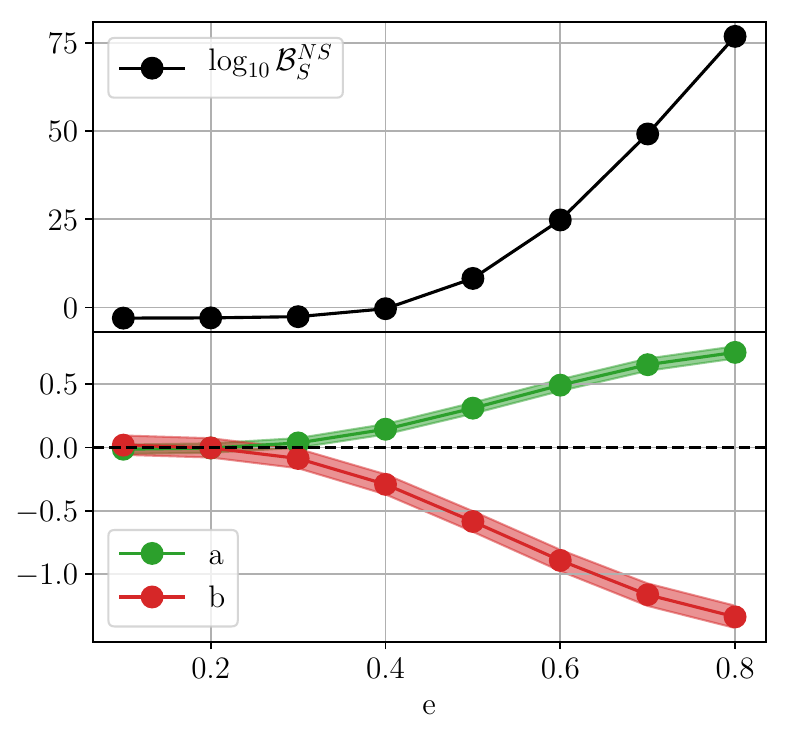}
	\caption{(top) Bayes factor for non-stationary against stationary GWB models as a function of the eccentricity of the injected source, (bottom) recovered posterior median of non-stationary parameters $a$ and $b$ as a function of eccentricity. The shaded region around $a$ and $b$ represents the $1-\sigma$ credible interval.}
	\label{fig:bf_ab_ecc}
\end{figure}
\autoref{fig:bf_ab_ecc}

\mf{The evolution of the recovered Bayes factor $\mathcal{B}^{NS}_{S}$ and parameters $a$ and $b$, as a function of the eccentricity $e$ of the injected GW signal, is illustrated in \autoref{fig:bf_ab_ecc}. It is evident that the model effectively captures the non-stationary features induced by the presence of the eccentric source when the latter is loud enough. The Bayes factor begins to favor the non-stationary model for eccentricities higher than 0.4. For this analysis, we have utilized the expression of $g(f,t)$ from \autoref{eq:nonstat_red_noise}. In future work, it might be beneficial to design a $g(f,t)$ that better suits this problem.}

\section{Discussion and conclusion}

The derived kernel serves the purpose of modeling non-stationarity in noise. It resembles the stationary kernels commonly used in PTA analyses but incorporates the capability to capture the time-dependent variance of the Gaussian weights. In its current form, this kernel remains an approximation and is valid only for small deviations from a stationary model. A significant challenge lies in understanding how correlations between frequencies can be accurately modeled, leading to an exact and non-approximated version of the non-stationary kernel. Further exploration in this area is necessary to improve our understanding and refine the methodology.

This approach necessitates a prior assumption about the expression of the evolutionary PSD. \mf{One significant advantage is its consistent expression of the covariance matrix for the entire dataset, unlike the time-slice analysis, which necessitates cutting the data into several segments.} This can be particularly valuable for applications like PTA, where noise is most pronounced in the lowest frequency bins.

\mf{In \autoref{sec:sim_results}, we showed that our model can occasionally produce broad posterior uncertainties for the non-stationary parameters, and the stationary solution may fall within the $1-\sigma$ credible interval. Consequently, accurately characterizing the evolutionary PSD can be challenging. Thus, it becomes crucial to assess the Bayes factor against the stationary model to determine whether the preference lies with a genuinely non-stationary signal.}

The slice analysis reveals a clear evolution in the inferred GWB properties. The first epoch appears much less informative than the following two. This can be attributed to an evolution of the data quality over time or the leakage of non-stationary features of individual pulsar noises into the GWB estimate. We conducted individual analyses for each pulsar to investigate non-stationary features and found that only J1713+0747 seemed to exhibit time-dependent DM noise amplitude variations. \mf{Non-stationary DM events were already reported for this pulsar in previous studies. Exponential dip events offer the best description of these events.}

The non-stationary model developed indicates that there is no evidence for a non-stationary GWB in the second data release of the EPTA data. Based on the results of the slice analysis, we infer that the apparent evolution of the inferred parameter values may stem from a poor characterization of individual pulsar noise contaminating the observed GWB. However, there is still a possibility that the presence of GW signals induced by eccentric SMBHBs in the data could produce similar non-stationary behaviors. \mf{The last section explored this hypothesis and showed that PTAs could detect non-stationary features of the GWB if an eccentric source is present in the data.}

With the current dataset, it remains inconclusive whether the observed evolution solely results from a change in backend quality or represents a genuine non-stationary feature of the GWB. Furthermore, if the noise from individual pulsars exhibits non-stationarity, it could potentially contaminate the GWB estimate, making it appear as non-stationary. Further investigation into this matter is warranted in future studies. Additionally, the upcoming release of the International Pulsar Timing Array (IPTA) third data release, with enhanced observation cadence and improved frequency coverage, is expected to provide better noise characterization and definitive insights into this issue.

\section*{Acknowledgements}

I thank Stas Babak, Philippa Cole, Aurélien Chalumeau and Hippolyte Quelquejay for the feedback on the manuscript. I also thank Philippe Bacon and Quentin Baghi for the interesting discussions.

M.F. acknowledges financial support  from MUR PRIN, Grant No.
2022-MYL2X, project 'GRAPE', funded by the European Union - Next
Generation EU. The Nan\c{c}ay radio Observatory is operated by the Paris Observatory, associated with the French Centre National de la Recherche Scientifique (CNRS), and partially supported by the Region Centre in France. IC, LG, GT acknowledge financial support from ``Programme National de Cosmologie and Galaxies'' (PNCG), and ``Programme National Hautes Energies'' (PNHE) funded by CNRS/INSU-IN2P3-INP, CEA and CNES, France. IC, LG, GT acknowledge financial support from Agence Nationale de la Recherche (ANR-18-CE31-0015), France. The Sardinia Radio Telescope (SRT) is funded by the Department of University and Research (MIUR), the Italian Space Agency (ASI), and the Autonomous Region of Sardinia (RAS) and is operated as a National Facility by the National Institute for Astrophysics (INAF). DP acknowledges support from the INAF Large Grant “Gravitational Wave Detection using Pulsar Timing Arrays” (P.I. D. Perrodin, 2023). JA acknowledges support from the European Commission (ARGOS-CDS; Grant Agreement number: 101094354). GMS sacknowledge financial support provided under the European Union’s H2020 ERC Consolidator Grant “Binary Massive Black Hole Astrophysics” (B Massive, Grant Agreement: 818691). The Westerbork Synthesis Radio Telescope is operated by the Netherlands Institute for Radio Astronomy (ASTRON) with support from the Netherlands Foundation for Scientific Research (NWO).

\section*{Appendix}

\subsection{2D spectral measure}
\label{sec:appendix_2dns}

The generalized Wiener-Khintchine relations are defined in \cite{priestley} as

\begin{equation}
\begin{aligned}
    C(t, t') & = \int df df' S(f, f') e^{2\pi i (t f - t' f')}\\
    S(f, f') & = \int dt dt' C(t, t') e^{-2\pi i (t f - t' f')}.
\end{aligned}
\label{eq:wk_2d}
\end{equation}

From the definition of the non-stationary covariance matrix in \autoref{eq:nonstat_kernel}, we can estimate the 2D spectral density:

\begin{equation}
\begin{aligned}
    S(f, f') = \int dt dt' \sum_i \sigma_i^2 g_i(t) g_i(t') & \bigg\{ e^{-2\pi i t(f+f_i)}e^{-2\pi i t'(-f-f_i)}\\
    & + e^{-2\pi i t(f_i-f)}e^{-2\pi i t'(f'-f_i)} \bigg\}\\
\end{aligned}
\end{equation}
with $\hat{g}_i(f)$ the Fourier transform of the function $g_i(t)$, we have

\begin{equation}
    S(f, f') = \sum_i \frac{\sigma_i^2}{2}\bigg[ \hat{g}_i(f+f_i)\hat{g}_i(-f'-f_i) + \hat{g}_i(-f + f_i)\hat{g}_i(f'-f_i)\bigg].
\end{equation}

Note that if $\hat{g}_i(f) = \delta_i(f)$ (a Dirac delta) and applying \autoref{eq:wk_2d}, we recover the stationary covariance matrix of \autoref{eq:covariance_stat}, giving a 2D spectral measure $S(f, f')$ that is only defined on the diagonal $f=f'$ and evaluated at discrete frequencies $f_i$. The spread of the function $\hat{g}_i (f)$ determines how much we deviate from the stationary covariance matrix.

\subsection{Simulations}
\label{sec:appendix_sim}

In section \ref{sec:gen_nonstat_rn}, we presented a method to generate a non-stationary colored noise. Here, we show an example of one realisation of non-stationary noise using this method. In \autoref{fig:corner_ns_rn}, we show the corner plots with the injected versus recovered values of parameter. We can see that the model performs well, except for parameter $b$, where the posterior uncertainties are quite large.

\begin{figure}[h]
	\centering
	\includegraphics[width=0.5\textwidth]{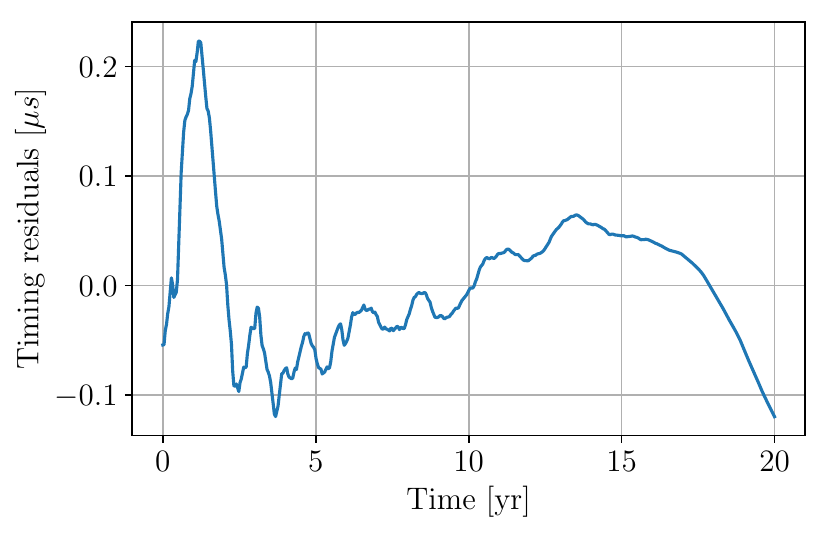}
	\caption{Time-domain plot of a non-stationary noise generated with the method presented in \ref{sec:gen_nonstat_rn}.}
	\label{fig:time_ns_rn}
\end{figure}

\begin{figure}[h]
	\centering
	\includegraphics[width=0.5\textwidth]{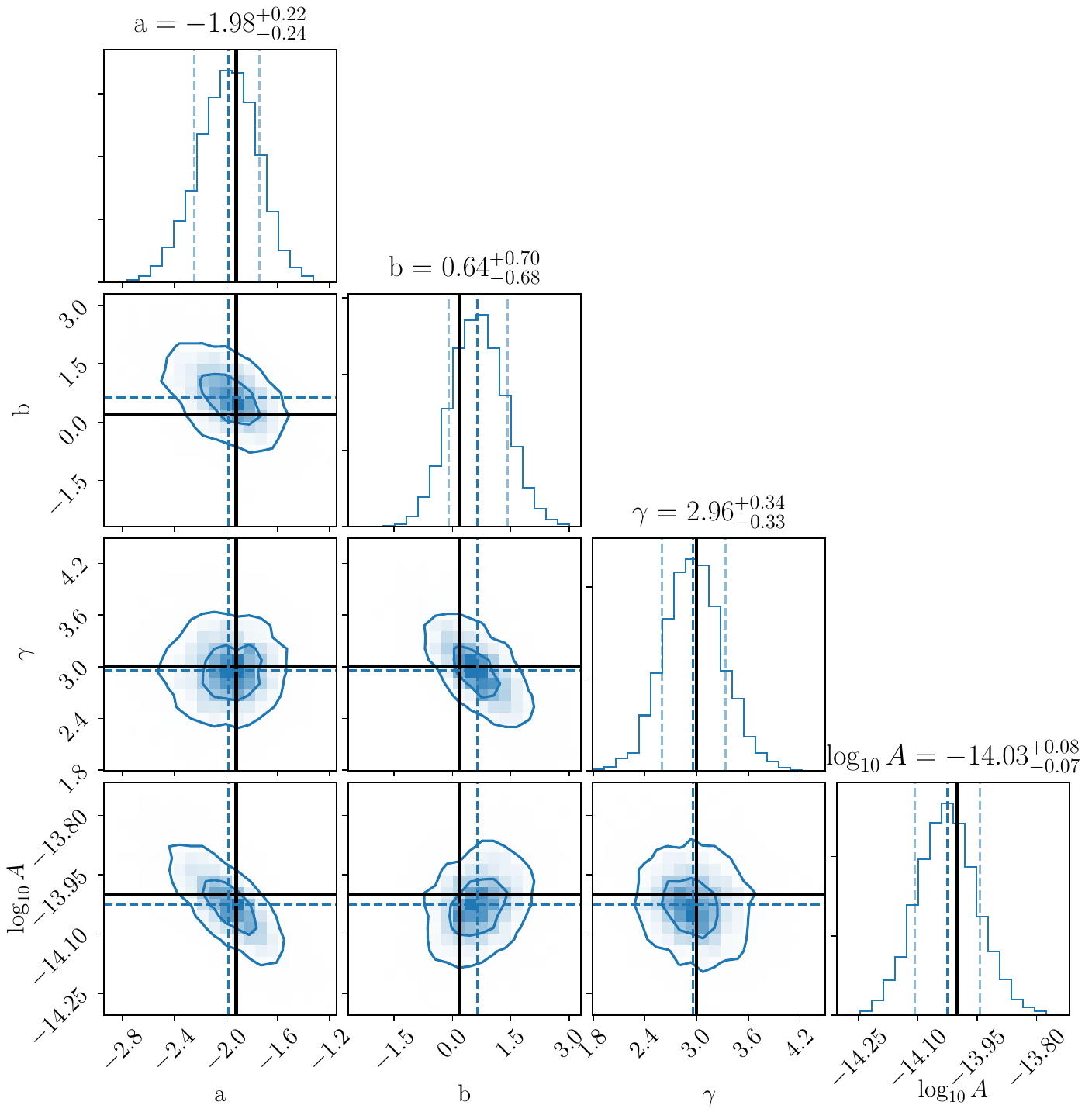}
	\caption{Corner plot of non-stationary parameters inferred from the simulation in \autoref{fig:time_ns_rn}. The solid black lines show the injected values of parameters. The contours on the 2d histograms are the $68\%$ and $95\%$ credible regions. The dashed lines on the 1d histograms show the $14/50/86\%$ percentiles.}
	\label{fig:corner_ns_rn}
\end{figure}

\subsection{PP-plots}

In \cite{cook}, a validation method is presented. It necessitates both data-generation and model-fitting software. In this context, our model-fitting software is PTMCMC \cite{justin_ellis_2017_1037579}, which has been previously utilized. During the fitting process, we sample the posterior distribution employing our model, specifically the non-stationary kernel, using the fitting software. In each simulation, we possess knowledge of the injected parameter values $\vec{\theta}_0$, allowing us to calculate the corresponding quantiles $q(\vec{\theta}_0)$ with respect to the posterior distribution derived from the fitting software. The methodology in \cite{cook} demonstrates that if the fitting software operates accurately and the generated data adheres to the same posterior distribution for parameters as the constructed model, then the quantiles $q(\vec{\theta}_0)$ must be distributed as Uniform(0, 1).

\begin{figure}[h]
	\centering
	\includegraphics[width=0.5\textwidth]{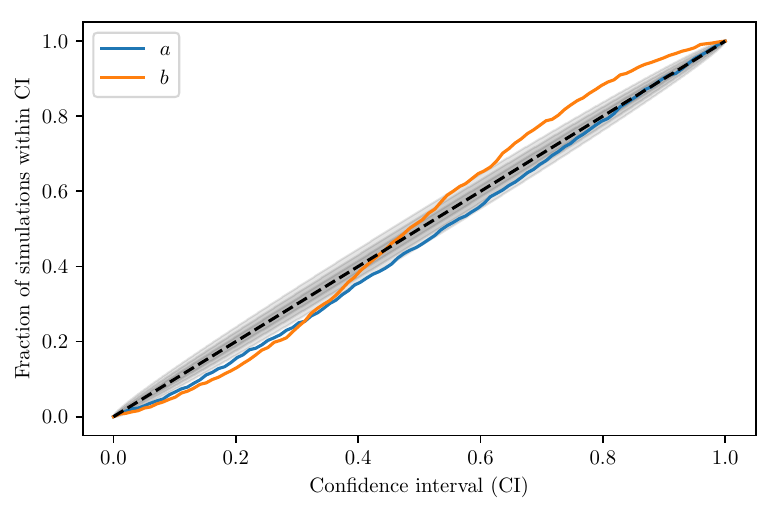}
	\caption{PP plot comparing the fraction of simulations within confidence interval (CI) and CI. The grey shaded region represent the $1-\sigma$ error from y=x.}
	\label{fig:pp_plot}
\end{figure}

In \autoref{fig:pp_plot} we show the PP plot for the quantiles $q(\vec{\theta}_0)$ obtained from simulations. They seem to follow the relationship y=x, but we notice discrepancies, especially for parameter $b$ which is the one with larger uncertainty.


\bibliography{bibliography.bib}

\end{document}